\documentclass[letterpaper,notitlepage]{article}
\usepackage{amsmath}

\input{tcilatex}

\begin{document}

\title{The volume of the black holes - the constant curvature slicing of the
spherically symmetric spacetime }
\author{Pawel Gusin*$^{1}$, Andrzej Radosz** \\
%EndAName
Wroc\l aw University of Science and Technology, Poland\\
*Faculty of Technology and Computer Science, \\
**Department of Quantum Technologies}
\maketitle

\begin{abstract}
We consider the problem of determination of a volume of some bounded
space-like hypersurfaces in the case of spherically symmetric spacetimes. In
the case when the hypersurfaces is cut or bounded by a light-like
hypersurface the problem may not be well defined. In order to define
properly the volume we introduce the volume forms related to the given
foliation (observer) of the considered spacetime. In the case of the
constant curvature slice the volume of the hypersurface cut by the horizon
(light-like surface) becomes composed of the two parts, outer and inner,
treated differently. We compute the corresponding volumes outside and inside
of the horizon of the ethernal Schwarzschild black hole.

\begin{description}
\item[PACS] 04.70.Bw ; 04.90.+e ; 04.20.Cv

\item[Key words] volume forms, observers
\end{description}
\end{abstract}

$^{1})$ {\small corresponding author: pawel.gusin@pwr.edu.pl}

\section{Introduction}

In order to find a volume of a space-like hypersurface $S$ of a spacetime $M$
one need to determine $S$ and to know a canonical volume form related to the
spacetime $M$. This canonical volume form is next pulled back to $S$ by an
embedding which gives the parametrization of $S$ and finally one performs an
appropriate integral. This procedure is well-known and well-defined. The
matter changes if one wants to find the volume of the space-like
hypersurface bounded or cut by an event horizon of a black hole, a
light-like hypersurface with vanishing volume but the fixed area. How to
define the foliation inside the horizon? The problem of foliations in the
case of the Schwarzschild spacetime appeared in [1-3]. In the papers [4-14]
the problem of a volume was considered for different types of black holes.
Different values of volumes had been found changing from: zero [8] to
infinity [13] being dependent on a chosen foliation $\mathcal{F}$\ of the
spacetime which gave the space-like slices of $M$. These slices are leaves
of $\mathcal{F}$.

Motivated by the above mentioned question: what is a volume of a spacelike
hypersurface $S$ cut by an event horizon we will consider in this paper a
volume as related to the constant curvature slice\ in the case of the
spherically symmetric spacetimes. We will demand that $S$ lays on the slice;
it is simultaneous with the slice. The condition of the constant curvature
of the slice is expressed by a constant value of the covariant divergency of
a vector $n$\ normal to $S$. Vanishing divergency of $n$ determines maximal
value of the volume of $S$. It results in this case in an equation that is
solved in order to determine the embeding of $S$ into $M$\ .The solution
contains some parameter $c$ of the dimension "length$^{2}$". Thus to find
the volume of $S\ $one needs to pull-back the volume form by this embedding
and to perform the integration in given limits. The embedding and also the
volume depend on $c$. This constant will be determined by the claim that the
induced volume functional reaches a local maximum with respect the
variations that keep fixed the boundary $\partial S$ of $S$.

In this paper we will consider the case of the spacetime that is
asymptotically Minkowskian, containing an ethernal black hole. In the
simplest case \ it is a given by Schwarzschild solution.

The paper is organized as follows:

In the section 2 we recall spherically symmetric metrics and we will obtain
the volume three-form for the foliation given by the unit time-like vector
(observer) in a case of a spherically symmetric spacetime. In the section 3
we will determine the embedding of the space-like hypersurface from the
condition of the constant mean extrinsic curvature. In the section 4 we will
apply this general approach to the case of the Schwarzschild black holes in
order to get a volume of hypersurface cut by the horizon and the volume of
the black holes in the given foliation (in three different asymptotics) The
section 5 is devoted to discussion. In the Appendix we prove that volume is
independent on the coordinate systems and we consider hypersurface bounded
by the trapped surfaces.

\section{Foliations and volume forms}

A spacetime $M$ is spherically symmetric if its isometry group contains a
subgroup isomorphic to the group $SO\left( 3\right) $, and orbits of this
subgroup are $2-$dimensional spheres $S^{2}$. In other words: the spacetime $%
M$ is spherically symmetric about one point $p$, if, in some coordinate
system, a metric $g_{\mu \nu }$ is invariant for three-dimensional spatial
rotations about $p$ that is, three-dimensional spatial rotations are
isometries for $g_{\mu \nu }$. However in the case of the Schwarzschild
spacetime $M$ the point $p$ is singular and not belongs to $M$. As is
well-known the metric $g_{\mu \nu }$ on $M$ can be expressed in the
coordinates $\left( x^{0},x^{1},\theta ,\phi \right) $ as follows:%
\begin{equation}
ds^{2}=g_{0}\left( x^{0},x^{1}\right) d\left( x^{0}\right) ^{2}-g_{1}\left(
x^{0},x^{1}\right) d\left( x^{1}\right) ^{2}-g_{2}\left( x^{0},x^{1}\right)
d\Omega _{S^{2}}^{2}  \tag{2.1}
\end{equation}%
and $d\Omega _{S^{2}}^{2}=d\theta ^{2}+\sin ^{2}\theta d\phi ^{2}$ is the
metric on the unit sphere $S^{2}$. In this space-time $M$\ the vector field $%
e_{\left( 0\right) }$ determines the orthonormal basis which is spanned by
vectors $e_{\left( a\right) }=X_{\left( a\right) }^{\mu }\partial _{\mu }$
such that $e_{\left( a\right) }\cdot e_{\left( b\right) }=X_{\left( a\right)
}^{\mu }X_{\left( b\right) }^{\nu }g_{\mu \nu }=\eta _{ab}=[diag\left(
+1,-1,-1,-1\right) ]_{ab}$ and $a,b=0,1,2,3$. We shall consider foliations
which have one basis vector fixed in the equatorial plane, $\theta =\pi /2$.
This vector is: $e_{\left( 2\right) }=\left( g_{2}\right) ^{-1/2}\partial
_{\theta }.$ Then, the coefficients $X_{\left( a\right) }^{\mu }$ are given
by the matrix (see [15]):

\begin{equation}
\left( X_{\left( a\right) }^{\mu }\right) =\left( 
\begin{array}{cccc}
\frac{\cosh q}{\sqrt{g_{0}}} & \frac{\sinh q\cos \chi }{\sqrt{g_{1}}} & 0 & 
\frac{\sinh q\sin \chi }{\sqrt{g_{3}}} \\ 
\frac{\sinh q}{\sqrt{g_{0}}} & \frac{\cosh q\cos \chi }{\sqrt{g_{1}}} & 0 & 
\frac{\cosh q\sin \chi }{\sqrt{g_{3}}} \\ 
0 & 0 & \frac{1}{\sqrt{g_{2}}} & 0 \\ 
0 & -\frac{\sin \chi }{\sqrt{g_{1}}} & 0 & \frac{\cos \chi }{\sqrt{g_{3}}}%
\end{array}%
\right) ,  \tag{2.2}
\end{equation}%
where $g_{3}=g_{2}\left( x^{0},x^{1}\right) \sin ^{2}\theta $. Thus the
foliation is parametrized by two functions $q$ and $\chi $ which depend on $%
x^{0}$ and $x^{1}$. The tetrad of this foliation is given by the four
one-forms $E^{\left( a\right) }=E_{\mu }^{\left( a\right) }dx^{\mu }$ dual
to the vector fields $e_{\left( a\right) }.$ It means that: $e_{\left(
a\right) }\cdot E^{\left( b\right) }=\delta _{a}^{b}$. The coefficients $%
E_{\mu }^{\left( a\right) }$ are given \ by the relation:%
\begin{equation}
\left( E_{\mu }^{\left( a\right) }\right) =\left[ \left( X_{\left( a\right)
}^{\mu }\right) ^{-1}\right] ^{T}.  \tag{2.3}
\end{equation}%
Thus:%
\begin{equation}
\left( E_{\mu }^{\left( a\right) }\right) =\left( 
\begin{array}{cccc}
\sqrt{g_{0}}\cosh q & -\sqrt{g_{1}}u_{1} & 0 & -\sqrt{g_{3}}u_{2} \\ 
-\sqrt{g_{0}}\sinh q & \sqrt{g_{1}}s_{1} & 0 & \sqrt{g_{3}}s_{2} \\ 
0 & 0 & \sqrt{g_{2}} & 0 \\ 
0 & -\sqrt{g_{1}}\sin \chi & 0 & \sqrt{g_{3}}\cos \chi%
\end{array}%
\right) ,  \tag{2.4}
\end{equation}%
where: $u_{1}=\sinh q\cos \chi $, $u_{2}=\sinh q\sin \chi $, $s_{1}=\cosh
q\cos \chi $ and $s_{2}=\cosh q\sin \chi $. In this tetrad $\left( E^{\left(
a\right) }\right) $ the metric has the form:%
\begin{equation}
ds^{2}=\left( E^{\left( 0\right) }\right) ^{2}-\left( E^{\left( 1\right)
}\right) ^{2}-\left( E^{\left( 2\right) }\right) ^{2}-\left( E^{\left(
3\right) }\right) ^{2}  \tag{2.5}
\end{equation}%
and the canonical volume four-form $dV$ is:%
\begin{gather}
dV=E^{\left( 0\right) }\wedge E^{\left( 1\right) }\wedge E^{\left( 2\right)
}\wedge E^{\left( 3\right) }=  \notag \\
=\det \left( E_{\mu }^{\left( a\right) }\right) dx^{0}\wedge dx^{1}\wedge
d\theta \wedge d\phi .  \tag{2.6}
\end{gather}%
Hence one finds that a volume three-form $\eta $ related to the foliation
determined by $e_{\left( 0\right) }$ is given by the inner product of $%
e_{\left( 0\right) }$ and $dV$ :%
\begin{equation}
\eta \equiv i_{e_{\left( 0\right) }}dV=E^{\left( 1\right) }\wedge E^{\left(
2\right) }\wedge E^{\left( 3\right) }.  \tag{2.7}
\end{equation}%
This form $\eta $ expressed in the coordinates $\left( x^{0},x^{1},\theta
,\phi \right) $ is:%
\begin{gather}
\eta =-g_{2}\sqrt{g_{0}g_{1}}\left( \frac{\cosh q}{\sqrt{g_{0}}}dx^{1}-\frac{%
\sinh q\cos \chi }{\sqrt{g_{1}}}dx^{0}\right) \wedge d\left( \cos \theta
\right) \wedge d\phi  \notag \\
-\sqrt{g_{0}g_{1}g_{2}}\sinh q\sin \chi dx^{0}\wedge dx^{1}\wedge d\theta , 
\tag{2.8}
\end{gather}%
where we used relation: $\sqrt{-\deg \left( g_{\mu \nu }\right) }=g_{2}\sqrt{%
g_{0}g_{1}}\sin \theta $. One can see that $\eta $ consists of two parts
that depend on $q$ and $\chi $. In the considered case of the spherical
symmetry $q$\ and $\chi $\ depend only on coordinates which appear in the
metric coefficients.\ 

\subsection{Spacetimes with black hole}

As is well-known a spacetime $\mathcal{M}$ with a non rotating and uncharged
black hole is the spherically symmetric solution of Einstein equations in
the vacuum with the cosmological constant $\Lambda $. In the
Schwarzschild-like coordinates ($x^{0}=T$, $x^{1}=R$) the solution is given
by metric: 
\begin{equation}
ds^{2}=h\left( R\right) dT^{2}-\frac{dR^{2}}{h\left( R\right) }-R^{2}\left(
d\theta ^{2}+\sin ^{2}\theta d\phi ^{2}\right) ,  \tag{2.12}
\end{equation}%
where 
\begin{equation}
h\left( R\right) =1-\frac{r_{S}}{R}-\frac{1}{3}\Lambda R^{2}  \tag{2.12a}
\end{equation}%
and $R\in (r_{S},R_{\max })$ the radius $R_{\max }$ depends on $r_{S}$ and $%
\Lambda $. The volume form $\eta $ in these coordinates takes the form:%
\begin{gather}
\eta =-\frac{R^{2}}{\sqrt{h\left( R\right) }}\left( \cosh qdR-h\left(
R\right) \sinh q\cos \chi dT\right) \wedge d\left( \cos \theta \right)
\wedge d\phi  \notag \\
-R\sinh q\sin \chi dT\wedge dR\wedge d\theta .  \tag{2.13}
\end{gather}%
Hereafter we will discuss the case $\Lambda =0$.

\section{Volume of the space-like slices}

A hypersurface $S$ in the spacetime $M$ is called space-like if a normal
vector field $n$ to $S$ is time-like. In our signature it means that $%
n^{2}>0 $. Such a hypersurface $S$ can be represented by means of its
embedding $\Phi :S\rightarrow M$ into the spacetime $M$\ via some parametric
equations:%
\begin{equation}
\Phi \left( \xi ^{i}\right) =\left( x^{\mu }\left( \xi ^{i}\right) \right)
\in M,  \tag{3.1}
\end{equation}%
where $\xi ^{i}$ are local coordinates for $S$ ($i=1,2,3$), while $x^{\mu }$
are local coordinates in $M$ ($\mu =0,...,3$).

In $M$ there is a canonical volume form $\varepsilon =\sqrt{\left\vert \det
\left( g_{\mu \nu }\right) \right\vert }dx^{0}\wedge dx^{1}\wedge d\theta
\wedge d\phi $. In order to get the volume of the hypersurface $S$ one need
to pull back the inner product $\varepsilon $ and $n$ by the embedding $\Phi 
$ and to perform an integral:%
\begin{equation}
vol\left( S;i_{n}\varepsilon \right) =\int_{S}\Phi ^{\ast }\left(
i_{n}\varepsilon \right) .  \tag{3.2}
\end{equation}%
An induced metric $h_{ij}$ on $S$ is given by the embedding $\Phi $ and the
metric $g_{\mu \nu }$ :%
\begin{equation}
h_{ij}=\frac{\partial x^{\mu }}{\partial \xi ^{i}}\frac{\partial x^{\nu }}{%
\partial \xi ^{j}}g_{\mu \nu }.  \tag{3.3}
\end{equation}%
Hence the volume form $\Phi ^{\ast }\left( i_{n}\varepsilon \right) $ reads:%
\begin{equation}
\Phi ^{\ast }\left( i_{n}\varepsilon \right) =\sqrt{\left\vert \det \left(
h_{ij}\right) \right\vert }d\xi ^{1}\wedge d\xi ^{2}\wedge d\xi ^{3}. 
\tag{3.4}
\end{equation}%
We shall consider the spherically symmetric space-like hypersurface $S$
embedded in the spherically symmetric spacetime $M$ with the metric (2.1) so 
$\Phi $ is:%
\begin{equation}
\Phi \left( \xi ,\theta ,\phi \right) =\left( x^{0}\left( \xi \right)
,x^{1}\left( \xi \right) ,\theta ,\phi \right)  \tag{3.5}
\end{equation}%
where $\xi \in \left[ \xi _{i},\xi _{f}\right] $ and $\theta ,\phi \in
S^{2}. $ The unit tangent vectors (in the metric (2.1)) to this space-like
hypersurface are:%
\begin{eqnarray}
t_{\left( 1\right) } &=&\frac{1}{\sqrt{g_{1}R^{\prime 2}-g_{0}T^{\prime 2}}}%
\left( T^{\prime }\frac{\partial }{\partial T}+R^{\prime }\frac{\partial }{%
\partial R}\right) ,  \notag \\
t_{\left( 2\right) } &=&\frac{1}{\sqrt{g_{2}}}\frac{\partial }{\partial
\theta },\text{ \ }t_{\left( 3\right) }=\frac{1}{\sqrt{g_{2}}\sin \theta }%
\frac{\partial }{\partial \phi }  \TCItag{3.6}
\end{eqnarray}%
and $t_{\left( a\right) }\cdot t_{\left( b\right) }=-\delta _{ab}$, where $%
a,b=1,2,3$ and the prime means differentiation with respect to $\xi :$ $%
T^{\prime }=\frac{dT}{d\xi },$ $R^{\prime }=\frac{dR}{d\xi }$ (and $%
x^{0}\equiv T,$ $x^{1}\equiv R,.$see Sec.2.1) Then the embedding $\Phi $ is
the space-like if 
\begin{equation}
g_{1}R^{\prime 2}>g_{0}T^{\prime 2}.  \tag{3.7}
\end{equation}%
The unit time-like vector $n$\ normal to $S$ is:%
\begin{equation}
n=\pm \frac{1}{\sqrt{g_{0}g_{1}\left( g_{1}R^{\prime 2}-g_{0}T^{\prime
2}\right) }}\left( g_{1}R^{\prime }\frac{\partial }{\partial T}%
-g_{0}T^{\prime }\frac{\partial }{\partial R}\right)  \tag{3.8}
\end{equation}%
where $n^{2}=+1$. The induced metric $h_{ij}$ on $S$ has the components:%
\begin{align}
h_{11}& =g_{0}T^{\prime 2}-g_{1}R^{\prime 2}\equiv -h_{1}<0,  \notag \\
h_{22}& =-g_{2},\text{ \ \ }h_{33}=-g_{2}\sin ^{2}\theta .  \tag{3.9}
\end{align}%
Thus we obtain that the volume of $S$ related to $\varepsilon $ is given by
the formula:%
\begin{equation}
vol\left( S;i_{n}\varepsilon \right) =4\pi \int_{\xi _{i}}^{\xi
_{f}}g_{2}\left( T\left( \xi \right) ,R\left( \xi \right) \right) \sqrt{%
g_{1}R^{\prime 2}-g_{0}T^{\prime 2}}d\xi .  \tag{3.10}
\end{equation}

On the other hand there is the volume form $\eta $ (2.8) related to an
arbitrary foliation which under the embedding $\Phi $ (3.5) takes the form:%
\begin{equation}
\Phi ^{\ast }\eta =-g_{2}\sqrt{g_{0}}\cosh q\left( \sqrt{\frac{g_{1}}{g_{0}}}%
R^{\prime }-T^{\prime }\tanh q\cos \chi \right) d\xi \wedge d\left( \cos
\theta \right) \wedge d\phi .  \tag{3.11}
\end{equation}%
Hence the volume of $S$ related to such foliation is:%
\begin{equation}
vol\left( S;\eta \right) =4\pi \int_{\xi _{i}}^{\xi _{f}}g_{2}\sqrt{g_{0}}%
\cosh q\left( \sqrt{\frac{g_{1}}{g_{0}}}R^{\prime }-T^{\prime }\tanh q\cos
\chi \right) d\xi .  \tag{3.12}
\end{equation}

The scalar product of the normal vector $n$ to $S$ and the unit time-like
vector $e_{\left( 0\right) }$ is:%
\begin{equation}
n\cdot e_{\left( 0\right) }=\frac{\cosh q}{\sqrt{h_{1}}}\left( \sqrt{g_{1}}%
R^{\prime }-\sqrt{g_{0}}T^{\prime }\tanh q\cos \chi \right) .  \tag{3.13}
\end{equation}%
One can say that $S$ lies on the simultaneity surface of the foliation given
by $e_{\left( 0\right) }$ if:%
\begin{equation}
n\cdot e_{\left( 0\right) }=1.  \tag{3.14a}
\end{equation}%
Hence we get the condition:%
\begin{equation}
\left( T^{\prime }-R^{\prime }\sqrt{\frac{g_{1}}{g_{0}}}\frac{\cosh q\sinh
q\cos \chi }{1+\sinh ^{2}q\cos ^{2}\chi }\right) ^{2}+\frac{g_{1}R^{\prime
2}\sinh ^{2}q\sin ^{2}\chi }{g_{0}\left( 1+\sinh ^{2}q\cos ^{2}\chi \right)
^{2}}=0.  \tag{3.14b}
\end{equation}%
which has the solution only for $\chi =0$ (since $g_{1}/g_{0}>0$) and: 
\begin{equation}
T^{\prime }=R^{\prime }\sqrt{\frac{g_{1}}{g_{0}}}\tanh q\left( R\left( \xi
\right) \right) .  \tag{3.15}
\end{equation}%
The embedding $\Phi $ takes the form: 
\begin{equation}
\Phi \left( R,\theta ,\phi \right) =\left( \int R^{\prime }\sqrt{\frac{%
g_{1}\left( R\right) }{g_{0}\left( R\right) }}\tanh q\left( R\left( \xi
\right) \right) d\xi ,R,\theta ,\phi \right) =\left( T\left( R\right)
,R,\theta ,\phi \right) ,  \tag{3.16}
\end{equation}%
and $T\left( R\right) $ is the solution of (3.15) expressed by the
coordinate $R$. The normal vector (3.8) for such an embedding is:%
\begin{equation}
n=\frac{\cosh q}{\sqrt{g_{0}}}\frac{\partial }{\partial T}-\frac{\sinh q}{%
\sqrt{g_{1}}}\frac{\partial }{\partial R}  \tag{3.17}
\end{equation}%
and the volume of $S$ is equal to:%
\begin{equation}
vol\left( S\right) =vol\left( S;i_{n}\varepsilon \right) =vol\left( S;\eta
\right) =4\pi \int_{R_{i}}^{R_{f}}\frac{g_{2}\sqrt{g_{1}}}{\cosh q}dR. 
\tag{3.18}
\end{equation}%
Hence the volume of $S$\ depends on the function $q.$\ But the foliation is
determined by $q$. As it is well known not every foliation is consistent
with the Einstein equations. In the ADM formalism the consistency conditions
are given by the constraint equations. However the dynamic equations give
the relation between the volume $vol\left( \mathcal{V}\right) $\ of a some
three-dimensional domain $V$\ with a boundary $\partial V$\ on the leaf $%
\Sigma _{t}$\ of the given foliation and an extrinsic curvature $K_{ij}$\ of 
$\Sigma _{t}$\ (eg. [16]) as follows:%
\begin{equation*}
\frac{dvol\left( \mathcal{V}\right) }{dn}=\int_{\mathcal{V}%
}h^{ij}K_{ij}d^{3}v,
\end{equation*}%
where $d^{3}v$\ is the volume form on $\Sigma _{t}$\ and $dvol\left( 
\mathcal{V}\right) /dn$\ means the change of the volume with the respect to
the normal vector $n$\ to $\Sigma _{t}$. One can say that the volume $%
vol\left( \mathcal{V}\right) $\ enclosed in $V$\ is extremal with respect to
variations of the domain delimited by $\partial V$, if the mean extrinsic
curvature $K\equiv h^{ij}K_{ij}$\ is vanishing. It is called maximal slicing
condition. We will apply this condition in our case and extend it to the
case when $K$\ takes the constant value. The mean extrinsic curvature $K$\
has a dimension of length$^{-1}$\ and is expressed by the vector $n$\ (see
eq. (3.8)) as follows: 
\begin{equation}
K=-\nabla _{\mu }n^{\mu }=\frac{-1}{\sqrt{-\det \left( g_{\alpha \beta
}\right) }}\partial _{\mu }\left[ \sqrt{-\det \left( g_{\alpha \beta
}\right) }n^{\mu }\right] .  \tag{3.19}
\end{equation}%
Thus if $K$\ is constant, then the function $q\left( T,R\right) $\ can be
determined from (3.19) in the special cases. If the vector field $n$\ were
the time-like Killing vector field, then one would obtain the thermodynamic
volumes [6,7,10,12]. But in our case $n$\ does not need to be the Killing
vector (see Sec. 4.2).

\section{Volume outside and inside horizon of the black hole}

In this section we will consider applications of this hypersurface volume
maximal value as given by means of the vanishing mean extrinsic curvature, $%
K=0$ in the case of the exterior and interior of the Schwarzschild black
hole. In both cases one demands that the hypersurface $S$ is space-like,
i.e. the condition (3.14a) is satisfied. Hence the volume of $S$ is given by
(3.12).

\subsection{The Schwarzschild black hole in Minkowski space-time}

\textbf{1) Outside horizon }

Let us start from an exterior of the Schwarzschild black hole in the
asymptotically flat Minkowski space-time. Thus in the metric (2.12) we put $%
\Lambda =0$. In this case, when the metric (2.12) expressed in Schwarzschild
coordinates, depends on $r$ but not on $t$, one can\ apply a conventional
notation,\ $T=t$, and $R=r$. Hence the space-like hypersurface $S$ may be
characterized as:%
\begin{equation}
S=\left\{ t=t\left( r\right) ,\text{ }r\in \left[ r_{1},r_{2}\right] \text{
and }\left( \theta ,\phi \right) \in S^{2}\right\} \subset \mathcal{M}. 
\tag{4.1}
\end{equation}%
It is the three dimensional manifold bounded by two spheres $S_{1}^{2}$ and $%
S_{2}^{2}$ with radii $r_{1}$ and $r_{2}$. Thus the volume (3.18) of $S$
related to the foliation given by the function $q$ is:%
\begin{equation}
vol\left( S\right) =4\pi \int_{r_{1}}^{r_{2}}\frac{g_{2}\left( r\right) 
\sqrt{g_{1}\left( r\right) }}{\cosh q\left( r\right) }dr=4\pi
\int_{r_{1}}^{r_{2}}r^{2}\sqrt{\frac{r}{r-r_{S}}}\frac{dr}{\cosh q\left(
r\right) },  \tag{4.2}
\end{equation}%
for $r_{2}>r_{1}>r_{S}$. Then the Eq. (3.19) takes the form :%
\begin{equation*}
K=\frac{1}{g_{2}\sqrt{g_{0}g_{1}}}\partial _{r}\left[ g_{2}\sqrt{g_{0}}\sinh
q\right]
\end{equation*}%
leading to the solution:%
\begin{equation}
\sinh q\left( r\right) =\frac{c}{g_{2}\sqrt{g_{0}}}+\frac{K}{g_{2}\sqrt{g_{0}%
}}\int^{r}g_{2}\sqrt{g_{0}g_{1}}dr^{\prime },  \tag{4.3}
\end{equation}%
where $c$ is a constant of dimension: length$^{2}$. One can see that the
function $q$ depends on two parameters $c$ and $K$ with dimensions:\ length$%
^{2}$ and length$^{-1}$, respectively. Hence the volume (3.18) of $S$ also
depends on these two parameters:%
\begin{equation}
vol\left( S;c,K\right) =4\pi \int_{r_{i}}^{r_{f}}\frac{g_{2}^{2}\left(
r\right) \sqrt{g_{0}g_{1}}}{\sqrt{g_{0}g_{2}^{2}+\left( c+K\int^{r}g_{2}%
\sqrt{g_{0}g_{1}}dr^{\prime }\right) ^{2}}}dr.  \tag{4.4}
\end{equation}%
The mean extrinsic curvature $K$ is the property of $S$ but the constant $c$
is arbitrary. However the volume (4.4) makes sense if its value is unique
and depends only on the geometry of the spacetime which is given by the
metric (2.1) and the geometry of $S$. If $K=0$, then%
\begin{equation}
\sinh q\left( r\right) =\frac{c}{g_{2}\sqrt{g_{0}}}  \tag{4.5a}
\end{equation}%
the volume (4.4)\ is:%
\begin{equation}
vol\left( S;c,0\right) =4\pi \int_{r_{i}}^{r_{f}}\frac{g_{2}^{2}\left(
r\right) \sqrt{g_{0}g_{1}}}{\sqrt{g_{0}g_{2}^{2}+c^{2}}}dr  \tag{4.5b}
\end{equation}%
and in this case it takes the form:%
\begin{equation}
vol\left( S;c,0\right) =4\pi \int_{r_{1}}^{r_{2}}\frac{r^{4}dr}{\sqrt{%
r^{4}-r^{3}r_{S}+c^{2}}}.  \tag{4.6}
\end{equation}%
Hence the volume of\ $S$, becomes the function of $c$. Introducing a
dimensionless variable $x=r/r_{S}$ the eq. (4.6) becomes:%
\begin{equation}
vol\left( S;C,0\right) =4\pi r_{S}^{3}\int_{r_{1}/r_{S}}^{r_{2}/r_{S}}\frac{%
x^{4}dx}{\sqrt{x^{4}-x^{3}+C^{2}}},  \tag{4.7}
\end{equation}%
where: 
\begin{equation}
C=c/r_{S}^{2}\geq 0.  \tag{4.8}
\end{equation}%
The expression (4.7) has been obtained in [13] for the interior of the event
horizon and led to the conclusion that the volume of the black hole is
infinite. In order to determine $C$ here we use the condition that in the
limit $r_{S}\rightarrow 0$ one reproduces the well-known result, volume in
the flat space. It means that (4.7) becomes%
\begin{equation}
vol_{0}\left( S;c,0\right) =4\pi \int_{r_{1}}^{r_{2}}\frac{r^{4}dr}{\sqrt{%
r^{4}+c^{2}}}  \tag{4.9}
\end{equation}%
(the index "$0$" labels\textbf{\ }the flat spacetime case). As the volume of 
$S$ in the flat space is:%
\begin{equation}
vol_{0}\left( S\right) =\frac{4\pi }{3}\left( r_{2}^{3}-r_{1}^{3}\right) . 
\tag{4.10}
\end{equation}%
Hence we obtain: 
\begin{equation}
c=c_{m}=0.  \tag{4.11}
\end{equation}%
\bigskip Let us underline that this value of $c$\ also extremises the volume
(4.7)\textbf{.}Therefore, Eq.(4.7) takes the form:%
\begin{equation}
vol\left( S;C=0\right) =4\pi r_{S}^{3}\int_{r_{1}/r_{S}}^{r_{2}/r_{S}}x^{2}%
\sqrt{\frac{x}{x-1}}dx.  \tag{4.12}
\end{equation}%
The above integral is elementary thus the volume of $S$ is:%
\begin{equation}
vol\left( S;C=0\right) =4\pi r_{S}^{3}\left[ I\left( x_{2}\right) -I\left(
x_{1}\right) \right] ,  \tag{4.13}
\end{equation}%
where $x_{1,2}=r_{1,2}/r_{S}$ and:%
\begin{equation}
I\left( x\right) =\frac{5}{8}\ln \left( \sqrt{x}+\sqrt{x-1}\right) +\sqrt{%
x\left( x-1\right) }\left( \frac{1}{3}x^{2}+\frac{5}{12}x+\frac{5}{8}\right)
,\text{ }x\geq 1.  \tag{4.14}
\end{equation}%
Expression (4.13) generalizes the meaning of the volume between the two
spheres. Indeed, in the limit of the flat spacetime, $r_{S}\rightarrow 0,$
(4.13) reduces to the flat space-time result (4.10).

One can notice that the function under the integral in the eq. (4.7) has the
asymptotic expansion:%
\begin{equation*}
\frac{x^{4}}{\sqrt{x^{4}-x^{3}+C^{2}}}\underset{x\rightarrow \infty }{\simeq 
}x^{2}+\frac{x}{2}+\frac{3}{8}
\end{equation*}%
which does not depend on the constant $C.$ Hence the volume between the two
spherical shells with the radii $r_{2}>r_{1}>>r_{S}$ is:%
\begin{equation*}
vol\left( S;c,0\right) =\frac{4\pi }{3}\left( r_{2}^{3}-r_{1}^{3}\right)
+\pi r_{S}\left( r_{2}^{2}-r_{1}^{2}\right) +\frac{3\pi }{2}r_{S}^{2}\left(
r_{2}-r_{1}\right) .
\end{equation*}

\textbf{2) Inside horizon}

In the case of the interior of the Schwarzschild black hole , $r<r_{S},$
radial $r$ and temporal, $t$ coordinates interchange their roles. Therefore
one can apply the following notation, $T=r$ and $R=t$. Thus the metric is:%
\begin{equation*}
ds^{2}=\frac{T}{r_{S}-T}dT^{2}-\left( \frac{r_{S}}{T}-1\right)
dR^{2}-T^{2}\left( d\theta ^{2}+\sin ^{2}\theta d\phi ^{2}\right)
\end{equation*}%
and has the components:

\begin{equation}
g_{0}\left( T\right) =g_{1}^{-1}\left( T\right) =\frac{1}{\frac{r_{S}}{T}-1},%
\text{\ }g_{2}\left( T\right) =T^{2}.  \tag{4.15}
\end{equation}%
Then the parameterizations (and the foliations) undertaken in sections 2 and
3 are left unchanged. We will consider the space-like hypersurface $S$
parametrized as follows:%
\begin{equation}
S=\left\{ T=T\left( \xi \right) ,\text{ }R=R\left( \xi \right) |\text{ }\xi
\in \left[ \xi _{1},\xi _{2}\right] \text{ and }\left( \theta ,\phi \right)
\in S^{2}\right\} \subset \mathcal{M}.  \tag{4.16}
\end{equation}%
The normal vector $n$ is given by (3.17) Thus the volume (3.18) of $S$\ is
extremized by the condition $K=0$:%
\begin{equation}
\partial _{T}\left[ g_{2}\sqrt{g_{1}}\cosh q\left( T\right) \right] =0. 
\tag{4.17}
\end{equation}%
And one finds that\ the function $q$\ is given by the relation:%
\begin{equation}
\cosh q\left( T\right) =\frac{c}{g_{2}\sqrt{g_{1}}}.  \tag{4.18}
\end{equation}%
In this case the condition (3.15) reads as follows:%
\begin{equation}
R^{\prime }=T^{\prime }\sqrt{\frac{g_{0}}{g_{1}}}\coth q\left( T\right) , 
\tag{4.19}
\end{equation}%
and the equation (3.18) becomes:%
\begin{equation}
vol\left( S;\eta \right) =4\pi \int_{T_{1}}^{T_{2}}\frac{g_{2}\left(
T\right) \sqrt{g_{0}\left( T\right) }}{\sinh q\left( T\right) }dT. 
\tag{4.20}
\end{equation}%
Using (4.18) we arrive to the equation:%
\begin{equation}
vol\left( S;C,0\right) =4\pi \int_{T_{1}}^{T_{2}}\frac{g_{2}^{2}\sqrt{%
g_{0}g_{1}}}{\sqrt{c^{2}-g_{1}g_{2}^{2}}}dT=4\pi
r_{S}^{3}\int_{T_{1}/r_{S}}^{T_{2}/r_{S}}\frac{x^{4}dx}{\sqrt{%
x^{4}-x^{3}+C^{2}}},  \tag{4.21}
\end{equation}%
where $C=c/r_{S}^{2}\geq 0$. In this range, $T<r_{S},$ the volume turns out
to be extremized by different values of parameter $c_{m}$ being dependent on
the range $\left[ T_{1},T_{2}\right] $. One can notice that the quadric
polynomial $w\left( x;C\right) =x^{4}-x^{3}+C^{2}$ has a minimum at $%
x_{m}=3/4$ and its value is: $w\left( 3/4;C\right) \equiv
w_{m}=C^{2}-3^{3}/2^{8}$. Thus $w\left( x;C\right) $ is grater or equal to
zero for all $x\geq 0$ if $C^{2}\geq C_{0}^{2}\equiv 3^{3}/2^{8}$. It means
that for $C>C_{0}$ the polynomial $w$ has no real roots. In the case when: $%
0<C<C_{0}$ there are two distinct positive roots $x_{1}$ and $x_{2},$
ordered as follows: $1>x_{2}>x_{1}.$ It means that $w>0$ for $x\in \lbrack
0,x_{1})\cup (x_{2},+\infty )$ and $w<0$ for $x\in \left( x_{1},x_{2}\right) 
$. If $C=0,$ then $w<0$ for $x\in \left( 0,1\right) $ and $w>0$ for $x\in
(1,+\infty )$. In the case when $C=C_{0}$ the polynomial $w$ has the
decomposition:%
\begin{equation}
w\left( x,C_{0}\right) =\left( x-\frac{3}{2^{2}}\right) ^{2}\left( x^{2}+%
\frac{1}{2}x+\frac{3}{2^{4}}\right) .  \tag{4.22}
\end{equation}
Thus the volume of the black hole $S_{BH}$ depends on $C:$%
\begin{equation}
vol\left( S_{BH};C,0\right) =4\pi r_{S}^{3}\int_{0}^{1}\frac{x^{4}dx}{\sqrt{%
x^{4}-x^{3}+C^{2}}}.  \tag{4.23}
\end{equation}%
Hence for $C^{2}=3^{3}/2^{8}$ the above integral is divergent and the volume
of the black hole is infinite (cf. [13]). Thus one can see that if $C=C_{0}$%
, then for $T_{2}=r_{S}$ and for $T_{1}>\frac{3}{4}r_{S}$ the volume is
finite but grows to infinity as $T_{1}$ approaches $\frac{3}{4}r_{S}$; then
it becomes infinite for\ $T_{1}<\frac{3}{4}r_{S}$.\emph{\ However the
regions between }$T_{1}=0$\emph{\ and }$T_{2}<\frac{3}{4}r_{S}$\emph{\ have
the finite volumes}\textit{.}

\subsection{The hypersurface cut by the horizon}

The obtained result may appear confusing. Above we have applied a unique
procedure leading to the determination of the volumes outside and inside
horizon. The set of formulae in both cases turned out to be different (cf.
Eqs. (4.2) , (4.20)) but the final outcomes were the same. Indeed the
volumes ouside horizon, Eq. (4.7) and the volume inside horizon,
Eq.(4.21).are expressed by the same integrals. One can ask then for the
volume\ of a spherically symmetric hypersurface $S$, $r_{1}<r_{S}<r_{2,}$
which is cut by the horizon.The volume of $S$ is then given formally by:

\begin{equation}
vol\left( S\right) =4\pi r_{S}^{3}\int_{r_{1}/r_{S}}^{r_{2}/r_{S}}\frac{%
x^{4}dx}{\sqrt{x^{4}-x^{3}+C^{2}}}.  \tag{4.24}
\end{equation}%
Assuming the continuity of the space outside and inside horizon (we consider
only the classical black hole) a choice $c=0$, might be made. Such an
approach leads however to the complex-value expression for the volume

\begin{eqnarray}
vol\left( S;c=0\right) &=&4\pi r_{S}^{3}\left( \int_{1}^{r_{2}/r_{S}}x^{2}%
\sqrt{\frac{x}{x-1}}dx+i\int_{r_{1}/r_{S}}^{1}y^{2}\sqrt{\frac{y}{1-y}}%
dy\right) =  \notag \\
&&4\pi r_{S}^{3}\left[ I\left( r_{2}/r_{S}\right) +i\left( \frac{5\pi }{16}%
-J\left( r_{1}/r_{S}\right) \right) \right] ,  \TCItag{4.25}
\end{eqnarray}%
where the functions $I$ (see 4.14) and $J$ are:%
\begin{equation}
I\left( x\right) =\frac{5}{8}\ln \left( \sqrt{x}+\sqrt{x-1}\right) +\sqrt{%
x\left( x-1\right) }\left( \frac{1}{3}x^{2}+\frac{5}{12}x+\frac{5}{8}\right)
,\text{ for }x\geq 1,  \tag{4.26}
\end{equation}%
\begin{equation}
J\left( y\right) =\frac{5}{8}\arctan \sqrt{\frac{y}{1-y}}-\sqrt{y\left(
1-y\right) }\left( \frac{1}{3}y^{2}+\frac{5}{12}y+\frac{5}{8}\right) ,\text{
for }0\leq y\leq 1.  \tag{4.27}
\end{equation}%
so that $I\left( 1\right) =J\left( 0\right) =0$ and $J\left( 1\right) =5\pi
/16.$ One can make a desperate step and\ to define the real volume of the $S$
as the modulus of $vol\left( S;c=0\right) $:%
\begin{equation}
|vol\left( S;c=0\right) |\equiv Vol\left( S;r_{1},r_{2}\right) =4\pi
r_{S}^{3}\left[ I^{2}\left( r_{2}/r_{S}\right) +\left( \frac{5\pi }{16}%
-J\left( r_{1}/r_{S}\right) \right) ^{2}\right] ^{1/2}.  \tag{4.28}
\end{equation}%
Hence the volume of the black hole given by the relation $\left(
r_{1}=0,r_{2}=r_{S}=2M\right) $ is:%
\begin{equation}
Vol\left( S_{BH}\right) =\frac{5}{4}\pi ^{2}r_{S}^{3}=10\pi ^{2}M^{3} 
\tag{4.29}
\end{equation}%
as opposed to the conclusion of [13] but this seems to be a very\textbf{\ }%
superficial attmept.

\subsection{The volume in the anti-de Sitter and de Sitter}

These considerations\ are easily extended to the case when $\Lambda \neq 0$
(anti-de Sitter or de Sitter). Thus (3.23) becomes:%
\begin{equation}
vol\left( S;0,K;\Lambda \right) =12\pi \int_{r_{i}}^{r_{f}}r^{2}\sqrt{\frac{r%
}{\left( K^{2}-9\Lambda \right) r^{3}+9r-9r_{S}}}dr.  \tag{4.30}
\end{equation}%
Here we also can take the constant $c=0$. As one finds there is a
hypersurface $S$ with a special value of $K$ in the de Sitter that the
volume of $S$ becomes the same as in the Minkowski spacetime One obtains
that if $K^{2}=9\Lambda $, then the hypersurface $S$ with $K^{2}=9\Lambda $
in the de Sitter has the same volume as the hypersurface $\widetilde{S}$
with $\widetilde{K}=0$ in Minkowski spacetime. The integral (4.30) can be
expressed by the elliptic integrals. However the final result is intricate
and does not bring new insights to the considered problem.

\section{Discussion}

The main motivation of this paper was the following question: what is the
volume of a spacelike spherically symmetric hypersurface $S$ cut by an event
horizon of the Schwarzschild black hole. This problem is obviously related
to the problem of the volume of a black hole recently discussed within
variety of approaches. Our proposal is to introduce different foliations
given by a two-parameter, $q,\chi ,$ velocity vector fields of a class of
observers. Such foliations are accompanied by the corresponding volume
forms, $\eta $ (2.8). On the other hand for a given spacelike hypersurface $%
S $ one can define an appropriate embedding $\Phi $ and related volume form
determining volume of $S$ (3.10). Another derivation of the\textbf{\ }%
expression for the volume of $S$ is given by using pull back of $\eta $ by $%
\Phi $ (3.12). Demanding that $S$ lays on the simultaneity surface (3.14a)
of the specific foliation one finds that these two volumes (3.10) and
(3.12), are equal.

The first interesting issue arises\textbf{\ }then as one finds, $\chi =0$.
In fact this corresponds to the claim of the spherical symmetry of the
foliation as the condition of vanishing $\chi $ represents the restoration
of the spherical symmetry of the world line $\gamma $ and simplifies the
volume form $\eta .$

One invokes then a maximum volume requirement: the volume of $S$ is maximal
if the mean extrinsic curvature $K$ vanishes. This is manifested differently
outside and inside horizon: geometry is spherically symmetric in both cases
but an interior being homogeneous along one spatial direction is dynamically
changing, whereas exterior is obviously static. In result one obtains
conditions different outside (4.3) and inside (4.18) horizon, both expressed
in terms of an unspecified constant $c$ (\textit{with dimension of length}$%
^{2}$) parameterizing the volume of $S$. Then the second interesting
observation may be made: although the condition $K=0$ for the exterior and
interior of the horizon is manifested differently, via (4.3) and (4.18), the
final expressions for the volume are the same in these two apparently
distinct regions and expressed in terms of variables $r$ and $T$ having
apparently different, spatial- and temporal-like, respectively, meaning, cf.%
\emph{\ }Eqs. (4.7) and (4.21).

There are two different ways to determine the parameter $c$ outside horizon:
apart of the demand to extremize the volume, one can chose $c$ in such a way
that in the flat spacetime limit, $r_{S}\rightarrow 0$, (Minkowski geometry)
the well-known result is reproduced. Both these ways lead to the same
result: $c_{m}=0$. As underlined above inside horizon the expression for a
volume of a spherically symmetric hypersurface $S$ is the same as the one
obtained outside horizon, but the extremising procedure provides $c_{m}$
being dependent on the boundaries of $S$, ($\xi _{1},\xi _{2}$ ),
corresponding to ($T_{1},T_{2}$) (see Sec. 4.1.2). So for the following
boundaries:

\begin{itemize}
\item $\frac{3}{2}M<T_{1}<T_{2}=2M$
\end{itemize}

\qquad\ $T_{1\text{ }}$decreases from $2M$ to $3/2M$, then $C_{m}\left(
T_{1}\right) $ increases from $0$ up to $C^{2}=C_{0}^{2}=3^{3}/2^{8}$; then
the volume grows indefinitely, i.e. it tends to infinity

\begin{itemize}
\item $T_{1}<\frac{3}{2}M<T_{2}=2M$

$T_{1\text{ }}$ further decreasing, from $3/2M$, to $0$, then $C_{m}\left(
T_{1}\right) $ becomes equal $C_{0}$ and the corresponding volume becomes
infinite;

\item $T_{1}<T_{2}=3/2M$
\end{itemize}

\qquad then $C_{m}$ becomes a function of $T_{2}$ and its value drops below $%
C_{0}$, then a finite volume in such a case is restored. (The dimensionless
parameter $C$\ has been defined as follows: $C=c/r_{S}^{2}$)

This leads to the conclusion concerning a volume of an ethernal black hole.
The volume of an ethernal Schwarzschild black hole is given by the maximal
value of the expression (4.21), attained for $%
C^{2}=C_{0}^{2}=3^{3}/2^{8}=27/256$ and turns out to be infinite -- hence
both the expression (4.21) as well as the following conclusions coincide
with the results derived by Christodulu and Rovelli [13]. However, two
important differences between above considerations and those of Ref. [13]
should be emphasized. First, the discussion presented in the Ref. [13] in
terms of Eddington-Filkenstein coordinates dealt with the case of a black
hole being formed due to the gravitational collapse and it was shown that in
fact the volume tends to infinity during this process. Second, the
maximizing condition was differently defined in Ref. [13] from the one given
here $K=0$, but the final expressions for the volume became identical.

Let us finally answer our initial the question about the volume of the
spherically symmetric hypersurface S, $r_{1}<r_{S}<r_{2}$ cut by the horizon
of the Schwarzschild black hole, $r=r_{S}=2M$. One finds that such a region
should be regarded as consisted of two distinct ranges: one, outside
horizon, ($r_{S},r_{2}$ ) and another one, inside horizon $\left(
r_{1}=T_{1},r_{S}=T_{2}\right) $. The volumes of these two ranges
parameterized with $c$ are determined via extremizing procedure with $%
c_{m}=0 $ and \ $c_{m}\left( T_{1}\right) $ (see above) in the former and in
the latter range, respectively. In this context the third interesting
outcome should be pointed out. As presented in Sec.4.2 in such a case one
can naively impose the flat spacetime condition $c_{m}=0$ for the whole
range $r_{1}<r_{S}<r_{2}$ obtaining however complex-valued volume.

One should remember however that $c$\ arises due to the requirement of
vanishing mean curvature, but on the other hand it represents the velocity
vector field, cf. Eqs. (4.3) and (2.2). The condition $c_{m}=0$\ defines
outside horizon an observer who is in the rest (cf. Eq (2.2)); such an
observer couldn't exists any longer inside horizon, hence one can not impose
such a demand for $r<r_{S}$. Similarly, appropriate demand $c_{m}\left(
T_{1}\right) $\ inside horizon (see above discussion) represents a condition
for an allowed observer (foliation) defined by a unit (time-like!) velocity
vector field\emph{.}

\section{Appendix}

\subsection{Transformation of the volume form}

The coordinates transformations which preserves the spherical symmetry have
the form:%
\begin{equation}
y^{\alpha }=y^{\alpha }\left( x^{A}\right) ,  \tag{a1}
\end{equation}%
where $\alpha =0,1$ and $A=0,1$. Under the above transformations the tetrads 
$E^{\left( a\right) }$ and $e_{\left( 0\right) }$ are:%
\begin{equation*}
E^{\left( a\right) }=E_{A}^{\left( a\right) }\frac{\partial x^{A}}{\partial
y^{\alpha }}dy^{\alpha }+E_{i}^{\left( a\right) }dx^{i},
\end{equation*}%
\begin{equation*}
e_{\left( 0\right) }=e_{\left( 0\right) }^{A}\frac{\partial y^{\alpha }}{%
\partial x^{A}}\frac{\partial }{\partial y^{\alpha }}+e_{\left( 0\right)
}^{i}\frac{\partial }{\partial x^{i}},
\end{equation*}%
where $x^{i}=\theta ,\phi $. Hence the volume form $\eta $ transforms as
follows:%
\begin{eqnarray}
\widetilde{\eta } &=&-g_{2}\sqrt{g_{0}g_{1}}\left( \frac{\cosh q}{\sqrt{g_{0}%
}}\frac{\partial x^{1}}{\partial y^{\alpha }}dy^{\alpha }-\frac{\sinh q\cos
\chi }{\sqrt{g_{1}}}\frac{\partial x^{0}}{\partial y^{\alpha }}dy^{\alpha
}\right) \wedge d\left( \cos \theta \right) \wedge d\phi  \notag \\
&&-\sqrt{g_{0}g_{1}g_{2}}\sinh q\sin \chi \det \left( \frac{\partial x^{A}}{%
\partial y^{\alpha }}\right) dy^{0}\wedge dy^{1}\wedge d\theta ,  \TCItag{a2}
\end{eqnarray}%
where the metric coefficients $g_{\mu }$ are functions of the new
coordinates $y^{\alpha }$. As one can see the three-form $\eta $ is
form-invariant under the transformations given by (a1).

Here we show that volume is invariant under the coordinate transformation.
Let in the coordinates $\left( t,r,\theta ,\phi \right) $ the hypersurface $%
S $ be given by the equation:%
\begin{equation*}
S=\left\{ \left( t,r,\theta ,\phi \right) |\text{ \ }\left( t\left( r\right)
,r,\theta ,\phi \right) \text{ \ and }r\in \left[ r_{1},r_{2}\right]
\right\} ,
\end{equation*}%
where $t\left( r\right) $ is given function. Then under the transformation:%
\begin{eqnarray}
T(t,r) &=&t+f\left( r\right) ,  \notag \\
R &=&r,  \TCItag{a3}
\end{eqnarray}%
$S$ transforms into $S_{\ast }$ which is given by :%
\begin{equation*}
S_{\ast }=\left\{ \left( T,r,\theta ,\phi \right) |\text{ \ }\left( t\left(
r\right) +f\left( r\right) ,r,\theta ,\phi \right) \text{ \ and }r\in \left[
r_{1},r_{2}\right] \right\} .
\end{equation*}%
Hence the pull-back of (a2) by $\widetilde{\Phi }$ is invariant:%
\begin{equation}
\widetilde{\Phi }^{\ast }\widetilde{\eta }=\Phi ^{\ast }\eta ,  \tag{a4}
\end{equation}%
where $\widetilde{\Phi }\left( r,\theta ,\phi \right) =$\ $\left( t\left(
r\right) +f\left( r\right) ,r,\theta ,\phi \right) $. Thus the volume of the
hypersurface is independent on the coordinates used in the computations. It
is also obvious that the condition (3.14a) is invariant under (a3). In this
way the results obtained in the section 3 remain valid in any coordinate
systems related by (a3).

\subsection{Hypersurfaces bounded by the trapped surfaces}

In the section 3 we considered hypersurface $S$ with the boundary $\partial
S $ which is sum of the two spheres and the volume of $S$ is given by (3.22)
for foliation with the constant mean extrinsic curvature. Here we express a
volume of a space-like hypersurface $S$ by the property of the boundary. It
is well know that the volume is given by:%
\begin{equation*}
vol\left( S\right) =\int_{\partial S}\sigma ,
\end{equation*}%
where $\sigma $ is a two-form, such that: $d\sigma =\Phi ^{\ast }\left(
i_{n}\varepsilon \right) =\Phi ^{\ast }\left( \eta \right) $. Thus:%
\begin{equation*}
vol\left( S\right) =\int_{M_{1}}\sigma -\int_{M_{2}}\sigma ,
\end{equation*}%
where $M_{1,2}$ are the space-like two-surfaces $\partial S=M_{1}\cup M_{2}$
and:%
\begin{equation*}
\sigma =F\left( r\right) ds_{1}\wedge ds_{2},
\end{equation*}%
$s_{1},s_{2}$ parametrize the boundary $\partial S$. The function $F\left(
r\right) $ is constant on $\partial S$. In the special case of the spherical
symmetry there are relations:%
\begin{equation*}
\frac{dF\left( r\right) }{dr}=\frac{g_{2}\left( r\right) \sqrt{g_{1}\left(
r\right) }}{\cosh q\left( r\right) }\text{ \ and }ds_{1}\wedge
ds_{2}=-d\left( \cos \theta \right) \wedge d\phi
\end{equation*}%
which leads to (3.18).

\section{References}

[1] B. L. Reinhart, \textit{Maximal foliations of extended Schwarzschild
space}, J. Math. Phys. 14 (1973) 719.

[2] S. Christensen, B. DeWitt, F. Estabrook, L. Smarr, E. Tsiang, and H.
Wahlquist, \textit{Maximally slicing a black hole}, Phys. Rev. D7 (1973)
2814.

[3] E. Malec, N. \'{O} Murchadha, \textit{Constant mean curvature slices in
the extended Schwarzschild solution and collapse of the lapse: Part I, }%
Phys.Rev. D68 (2003) 124019, [arXiv: gr-qc/0307046]

[4] M. Parikh, \textit{Volume of black holes}, Phys. Rev. D 73 (2006)
124021, [arXiv:hep-th/0508108].

[5] D. Grumiller, \textit{The Volume of 2D Black Holes}, J.Phys.Conf.Ser. 33
(2006) 361-366,\ [arXiv:gr-qc/0509077].

[6] W. Ballik and K. Lake, \textit{The volume of stationary black holes and
the meaning of the surface gravity},\ [arXiv:1005.1116].

[7] W. Ballik and K. Lake, \textit{Vector volume and black holes}, Phys.
Rev. D 88 (2013) 104038, [arXiv:1310.1935].

[8] B. S. DiNunno and R. A. Matzner, \textit{The volume inside a black hole}%
, General Relativity and Gravitation 42 (2009) 63--76, [arXiv:0801.1734].

[9] T.K. Finch, \textit{Coordinate families for the Schwarzschild geometry
based on radial timelike geodesics}, [arXiv:1211.4337].

[10] M. Cveti\v{c}, G. W. Gibbons, D. Kubi\v{n}\'{a}k, and C. N. Pope, 
\textit{Black hole enthalpy and an entropy inequality for the thermodynamic
volume},\ Phys. Rev. D 84 (2011) 024037, [arXiv:1012.2888].

[11] G. W. Gibbons, \textit{What is the Shape of a Black Hole? }AIP\textit{\ 
}Conf. Proc. 1460 (2012) 90--100,\textit{\ }[arXiv:1201.2340].

[12] B. P. Dolan, D. Kastor, D. Kubiz\v{n}\'{a}k, R. B. Mann, and J.
Traschen, \textit{Thermodynamic volumes and isoperimetric inequalities for
de Sitter} \textit{black holes,} Phys. Rev. D 87, (2013) 104017,\
[arXiv:1301.5926].

[13] M. Christodoulou, C. Rovelli,\textit{\ How big is a black hole?}, Phys.
Rev. D 91 (2015), 064046, [arXiv: 1411.2854].

[14] N. Altamirano, D. Kubiznak, R.B. Mann, Z. Sherkatghanad, \textit{%
Thermodynamics of rotating black holes and black rings: phase transitions
and thermodynamic volume}. Galaxies. 2(4), 89--159 (2014).
http://www.mdpi.com/2075-4434/2/1/89, [arXiv:1401.2586].

[15] P. Gusin, B. Ku\'{s}nierz, \ A. Radosz, \textit{Observers in spacetimes
with spherical and axial symmetries}, [arXiv:1507.01617].

[16] C.W. Misner, K.S. Thorne, and J.A. Wheeler, \textit{Gravitation},
Freeman, New York (1973).

\bigskip

\end{document}